\documentclass[aps,pre,twocolumn,superscriptaddress,showpacs,amsmath,amssymb]{revtex4-1}

\pdfoutput=1 

\usepackage{graphicx}
\usepackage{dcolumn}
\usepackage{subfigure}

\renewcommand{\Im}{\mathop{\mathrm{Im}}}

\begin{document}

\title{Conductance of DNA molecules: Effects of decoherence and bonding}

\author{Mat\'ias Zilly}
\affiliation{Department of Physics and CeNIDE, University of
  Duisburg-Essen, D-47048 Duisburg, Germany} 
\homepage{http://www.uni-due.de/comphys}
\author{Orsolya Ujs\'aghy}
\email{ujsaghy@neumann.phy.bme.hu}
 \affiliation{Department of Theoretical Physics and Condensed
   Matter Research Group of the Hungarian Academy of Sciences,\\
   Budapest University of Technology and Economics, Budafoki \'ut 8.,
   H-1521 Budapest, Hungary}
\author{Dietrich E.\ Wolf}
\affiliation{Department of Physics and CeNIDE, University of
  Duisburg-Essen, D-47048 Duisburg, Germany}

\date{\today}

\begin{abstract}
The influence of decoherence and bonding on the linear conductance of single
double-stranded DNA molecules is examined by fitting a
phenomenological statistical model developed recently (EPJB {\bf 68},
237 (2009)) to experimental results. The DNA molecule itself is
described by a tight binding ladder model with parameters obtained
from published ab initio calculations (J.Am.Chem.Soc. {\bf 127}, 14894
(2005)). The good agreement with the experiments on sequence and length
dependence gives a hint on the nature of conduction in DNA and at the
same time provides a crucial test of the model.
\end{abstract}


\maketitle

\section{Introduction\label{intro}}

Motivated by 
molecular electronics and by the possibility to read out
chemical and biological information electronically, transport
through single double stranded DNA molecules is a focus of current
research in nanoscience. The experimental situation is not
particularly clear. After the early investigations
\cite{*[{For a summary, see }] Endres-Cox-Singh_2004}, in the last six
years increasing consensus emerged  that DNA
double strands with 8 to 26 base pairs are conducting. 
However, this conclusion may be misleading, as it depends on the
chemical potential of the contacts, on the temperature, and on the coupling of
DNA to the environment. The experiments are performed either in water
\cite{Xu-Zhang-Li-Tao_2004,Hihath-Xu-Zhang-Tao_2005,Hihath-Chen-Zhang-Tao_2007}
or in the dry state
\cite{Mahapatro-Jeong-Lee-Janes_2007,Dulic-Tuukkanen-Chung-Isambert-Lavie-Filoramo_2009}.
Moreover, all experiments show sequence dependent conductance, which
increases the more the  
(GC)-content in the sample molecule dominates over the (AT)-content.
Sometimes ohmic behavior is reported,
sometimes one finds an exponential decrease which is
attributed to coherent tunneling. The change from coherent to
ohmic behavior is usually attributed to the effect of decoherence
caused by the water (e.g.
\cite{Gutierrez-Mandal-Cuniberti_2005,Mallajosyula-Lin-Cox-Pati-Singh_2008,
  Starikov-Quintilla-Ngabou-Lee-Cuniberti-Wenzel_2009}), vibrational
degrees of freedom
\cite{Schmidt-Hettler-Schon_2008,*Schmidt-Hettler-Schon_2007}, or
dynamical movement of the DNA bases
\cite{Gutierrez-Caetano-Woiczikowski-Kubar-Elstner-Cuniberti_2009}.
Most of the theoretical studies perform model calculations \cite{[{For
    a review of the tight-binding models used to describe transport in
    DNA, see }] Cuniberti_2007}, but some of them present hybrid
methods combining \textit{ab initio} and Molecular Dynamics studies
\cite{Mallajosyula-Lin-Cox-Pati-Singh_2008}, sometimes also in
combination with model calculations
\cite{Gutierrez-Caetano-Woiczikowski-Kubar-Elstner-Cuniberti_2009}.

The experiments clearly show that both the way, DNA is
bonded, as well as the environment have an influence on the conductance.
Here we present a calculation taking into account the effects of
decoherence and bonding by extending a
recently introduced phenomenological model \cite{ZUW_2009,MatiasPhD}.  
It was originally developed for linear systems, and adapting 
it to the quasi-linear transport in DNA provides additional 
justification of the model.

As DNA-Hamiltonian we use a realistic tight
binding model (the extended ladder model) based on \textit{ab initio}
calculations 
\cite{Senthil_2005,Cuniberti_Elstner_2008,Zhang_2002}. 

We investigate several sequences, for which experimental values for
the conductance are known, in order to assess, how well they can be
fitted by our model {\em without changing the microscopic energy
  values}, but just {\em adapting the four parameters} $\mu$ and 
$\Gamma$ describing the coupling to the electrodes, and $p$ and $\eta$
describing the effect of decoherence (see Sections \ref{bonding} and
\ref{decoh}). 

The paper is organized as follows. In Sect.~\ref{DNA} we present the
extended ladder model describing the double stranded DNA.
Sect.~\ref{bonding} is devoted to modelling the bonding to the
electrodes according to the the experimental situation. In
Sect.~\ref{decoh} we extend the statistical model for
decoherence such that it becomes applicable to DNA. 
In Sect.~\ref{exp} we compare our results to the experiments 
\cite{Xu-Zhang-Li-Tao_2004,Hihath-Xu-Zhang-Tao_2005,Hihath-Chen-Zhang-Tao_2007} and 
finally, we give our conclusions in Sect.~\ref{concl}.

\section{Extended ladder model of DNA\label{DNA}}

The linear chain tight binding model \cite{[{For a review of the
    tight-binding models used to describe transport in DNA, see }]
  Cuniberti_2007}, where each site
corresponds to a base pair, is widely used in the literature to
describe DNA. In some works also the backbone
effects caused by the complementary strand and the sugar/phosphate
mantle are taken into account (fishbone model) \cite{[{For a review of the
    tight-binding models used to describe transport in DNA, see }]
  Cuniberti_2007}. 

However, in order to account for arbitrary base sequences considering
the correct base pairing, the so called ladder model
\cite{[{For a review of the tight-binding models used to describe
    transport in DNA, see }] 
Cuniberti_2007,Wei-Chan_2007} is more
appropriate. It consists of two coupled tight binding chains
corresponding to the two strands, and each site represents a single
base. There are calculations where backbone effects (here only due to
the sugar/phosphate mantle) are considered in the ladder model, 
as well \cite{[{For a review of the
    tight-binding models used to describe transport in DNA, see }] Cuniberti_2007}.

\begin{figure}
\includegraphics[width=8truecm]{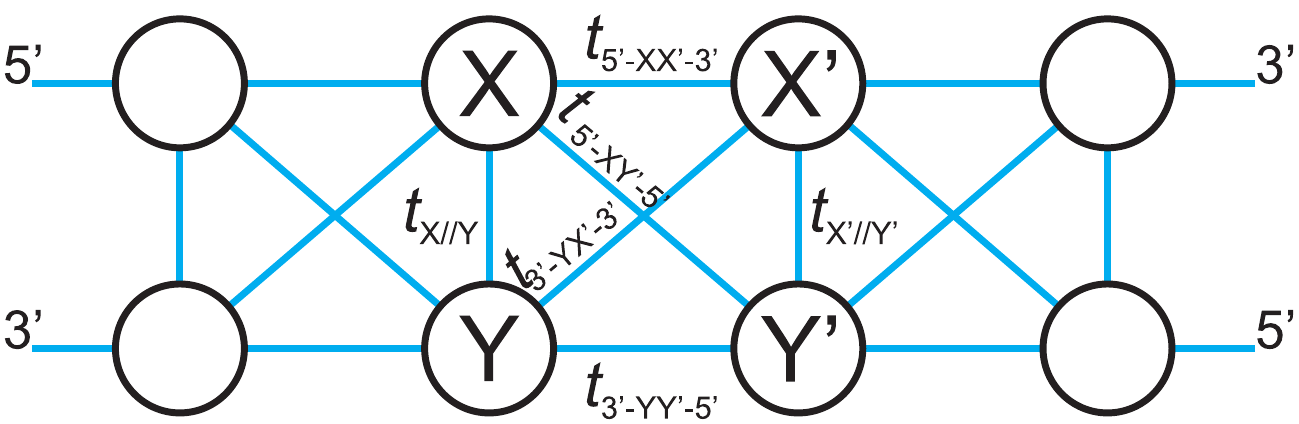}
\caption{The extended ladder model for DNA double strands 
\cite{Senthil_2005,Cuniberti_Elstner_2008,Zhang_2002}.\label{DNA_ext_model}}
\end{figure}

Since, according to \textit{ab-initio}
\cite{Senthil_2005,Cuniberti_Elstner_2008,Zhang_2002} calculations the diagonal
inter-strand transfer matrix elements can be more relevant than the
intra-strand coupling, we will use the extended ladder model of DNA 
\cite{Senthil_2005,Cuniberti_Elstner_2008,Zhang_2002} sketched 
in Fig.~\ref{DNA_ext_model}. 
The corresponding Hamiltonian describing a double-stranded DNA that is
$N$ base pairs long is
\begin{eqnarray}
  \label{DNA_Hamilton}
H&=&\sum\limits_{i=1}^{N} \bigl [\sum\limits_{m=1,2} \varepsilon_{i,m}
c^{\dagger}_{i,m} c_{i,m}  +\sum\limits_{m,n=1,2;m\neq n} t_{i,mn}
c^{\dagger}_{i,m} c_{i,n} \cr 
&+&\sum\limits_{m=1,2}  t_{i,i+1,mn} \bigl [c^{\dagger}_{i,m} c_{i+1,n} + \text{h.c.}\bigr ]
\bigl ],
\end{eqnarray}
where $c^{\dagger}_{i,m}$ creates a hole on
strand $m$ at the $i$th base pair with on-site energy $\varepsilon_{i,m}$. 
$t_{i,mn}$ and $t_{i,i+1,mn} $ are the base pair couplings and
the hopping amplitudes between two bases of neighboring base pairs,
respectively. The parameters used in the calculations for 
the on-site energies and the base pair couplings
are listed in Table~\ref{DNA_on-site_intrabasepair}. 
According to \cite{Senthil_2005}, the on-site energies calculated
by density functional theory (DFT) depend on the flanking nucleo\-bases,
so that 16 different values were determined. For $\epsilon_\text{G}$
they vary between 7.890eV and 8.407eV, to give an idea. Here, we
want to propose a universal model, which can predict conductances for
any sequence of base pairs. Within the model class
(\ref{DNA_Hamilton}) a single value for the on-site energy of a
nucleobase should be used. We simply take the average of the 16 values
given in \cite{Senthil_2005} for each nucleobase. 
 
\begin{table}[htbp]
  \begin{center}
   \begin{tabular}{cccc|cc}
      $\epsilon_\text{G}$ & $\epsilon_\text{A}$ & $\epsilon_\text{C}$
      & $\epsilon_\text{T}$ & $t_\text{G//C}$ & $t_\text{A//T}$ \\ \hline
      8.178 & 8.631 & 9.722 & 9.464 & -0.055 & -0.047
    \end{tabular}
\end{center}
\caption{The on-site energies and base pair
     couplings (in eV). On-site energies are averages of the values
     calculated by DFT
     \cite{Senthil_2005}. They
     correspond to the HOMO orbitals at the respective bases.
\label{DNA_on-site_intrabasepair}}
\end{table}
The hopping parameters (cf.~Fig.~\ref{DNA_ext_model}) used in the calculation
are listed in Table~\ref{DNA_parameters}. We use
the single-strand notation, listing only the sequence of a
single strand (the other strand is determined due to the unique base
pairing). Because of
the directionality of the DNA strands, $t_{\mbox{5'-XY-3'}}\neq
t_{\mbox{3'-XY-5'}}=t_{\mbox{5'-YX-3'}}$ for $\mbox{X}\neq\mbox{Y}$. However, due to symmetry
$t_{\mbox{5'-XY-5'}}=t_{\mbox{5'-YX-5'}}$ and $t_{\mbox{3'-XY-3'}}=t_{\mbox{3'-YX-3'}}$ for all X,Y.
\begin{table}[htbp]
\begin{center}
\subtable[$t_\text{5'-XY-3'}=t_\text{3'-YX-5'}$]{\label{tab:5-XY-3}
\begin{tabular}{c|cccc}
    X\textbackslash Y & G & A & C & T \\ \hline
    G & 0.053 & - 0.077 & -0.114 & 0.141 \\
    A & -0.010 & -0.004 & 0.042 & -0.063 \\
    C & 0.009 & -0.002 & 0.022 & -0.055 \\
    T & 0.018 & -0.031 & -0.028 & 0.180
  \end{tabular}}\\ 
 \subtable[$t_\text{5'-XY-5'}$]{\label{tab:5-XY-5}\begin{tabular}{c|cccc}
    X\textbackslash Y & G & A & C & T \\ \hline
    G & 0.012 &-0.013 & 0.002 & -0.009	\\
    A & -0.013 & 0.031 & -0.001	& 0.007 \\
    C & 0.002 & -0.001 & 0.001 & 0.0003 \\
    T & -0.009 & 0.007 & 0.0003 & 0.001
  \end{tabular}} \qquad 
  \subtable[$t_\text{3'-XY-3'}$]{\label{tab:3-XY-3}\begin{tabular}{c|cccc}
    X\textbackslash Y & G & A & C & T \\ \hline
    G & -0.032 & -0.011 & 0.022 & -0.014 \\
    A & -0.011 & 0.049 & 0.017 & -0.007	\\
    C & 0.022 & 0.017 & 0.010 & -0.004 \\
    T & -0.014 & -0.007 & -0.004 & 0.006 
  \end{tabular}}
\end{center} 
 \caption{Hopping parameters (in eV) for the extended ladder model,  
   Fig.~\ref{DNA_ext_model} \cite{Senthil_2005}. 
\label{DNA_parameters}}
\end{table}

\begin{figure}
\includegraphics[height=5truecm]{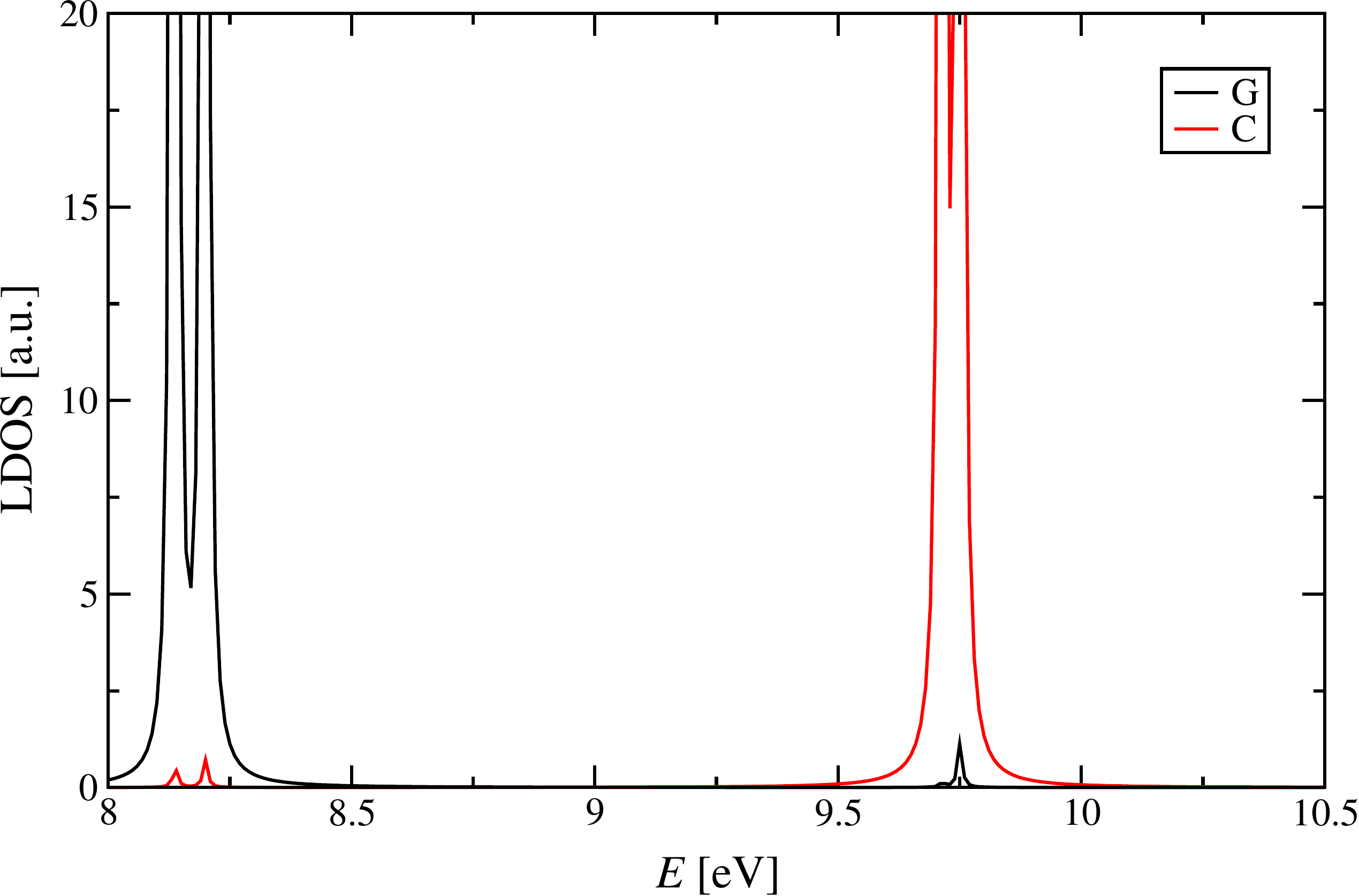}
\caption{Density of states for holes, for 
$\mbox{5'-(CG)}_{\infty}\mbox{-3'}$.
\label{inf_DOS}}
\end{figure}

Using the parameters given in Tables
~\ref{DNA_on-site_intrabasepair} and \ref{DNA_parameters}
the density of states for holes of an infinite DNA double strand
is shown in Fig.~\ref{inf_DOS}. We can see two bands, one for holes
with an energy around 8.2eV and one around 9.7eV. Each band is split
into two subbands of width $\approx 0.04$eV. We prefer to describe
the bands in terms of electrons in the following: The upper band at an
energy $-8.2$eV below the vacuum level and the lower one at $-9.7$eV
are both filled with electrons for an isolated neutral molecule.

\section{Bonding Model \label{bonding}}

\begin{figure}
\includegraphics[width=8.5truecm]{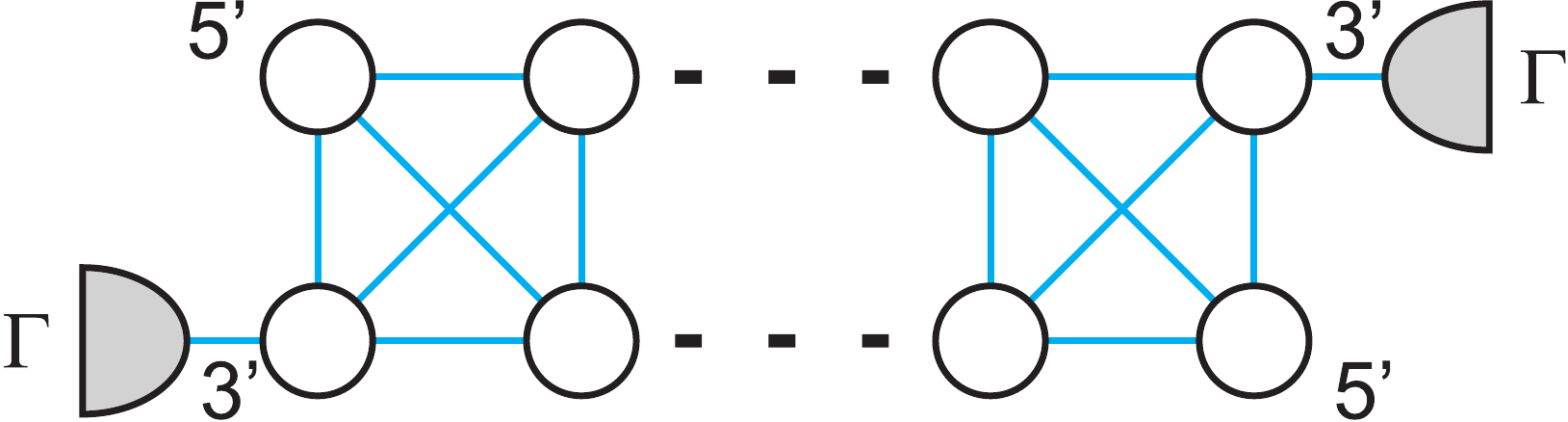}
\caption{Extended ladder model, where only the 3'-ends of the
  strands are attached to the gold contacts via thiol-linker like in
  the experiments of 
  \cite{Xu-Zhang-Li-Tao_2004,Hihath-Xu-Zhang-Tao_2005,
Hihath-Chen-Zhang-Tao_2007}.
\label{contact}}
\end{figure}

Our aim is to examine the effect of the environment (decoherence) on
the linear conductance of single double stranded DNA molecules. 
We focus on the experiments in 
\cite{Xu-Zhang-Li-Tao_2004,Hihath-Xu-Zhang-Tao_2005,
Hihath-Chen-Zhang-Tao_2007}, where each gold electrode is coupled to a
G-base on the 3'-end via a thiol group.
We model this as shown in Fig.~\ref{contact}, assuming 
wide-band limit contacts attached only to the the 3'-ends of the
strands. The self-energies due to the left (L) and right (R) contacts are
$2N\times 2N$ matrices, where all matrix elements are zero apart from
the diagonal elements corresponding to $(i,m)=(1, 2)$ on the left and
$(i,m)=(N, 1)$ on the right. These matrix elements are equal to the
imaginary number $-\mathrm{i}\Gamma/2$. 
We will consider the case of weak coupling to the electrodes, by
  choosing $\Gamma=0.003$eV. This is of the same order of magnitude
  as the weakest hopping parameters in the DNA molecule.

The conductance of the DNA fragment depends also on the Fermi level of
the contacts in relation to the DNA bands. Already a relative shift of
0.01eV can significantly change the conductance. As the known values are
hardly that reliable, we consider the chemical potential $\mu$ of the
contacts as a fit parameter in the following. It must be kept in
  mind that $\mu$  
  denotes the chemical potential for holes. The one for electrons is
  given by $\mu_{\rm el}=-\mu$. We concentrate on values of $\mu_{\rm
    el}$ within the immediate neighborhood of the uppermost occupied band,
i.e. around $-8.18$eV. Hence we describe
the bonding of the DNA fragment by two fit parameters, the self energy
$\Gamma$ and the chemical potential $\mu$.

In Sections \ref{decoh} and \ref{exp} we will calculate, how the linear
conductance of the 
experimentally investigated sequences depends on these two bonding
parameters. They should be the same for all sequences in   
\cite{Xu-Zhang-Li-Tao_2004,Hihath-Xu-Zhang-Tao_2005,Hihath-Chen-Zhang-Tao_2007},
because the bonding was done the same way in all samples.    
We are going to use the recently developed phenomenological statistical method
\cite{ZUW_2009,MatiasPhD}, which takes decoherence into account by two
further fit parameters. 

\section{Statistical model for the decoherence \label{decoh}}

The model is based on a different physical picture compared to other 
phenomenological descriptions of decoherence (like the method of fictitious reservoirs 
(B\"uttiker probes) \cite{Buttiker_1986,*Damato-Pastawski_1990} or the
method of \cite{Golizadeh-Mojarad-Datta_2007}). Whereas there
decoherence is continuously present in 
the sample, in our model it occurs only at stochastically
distributed decoherence regions, where phase information gets
completely lost. 

We assume that decoherence can be described by local, stochastic
events that couple a base pair (or part of it) to the environment. At
those positions, phase coherence will be lost in both strands due to
the interstrand couplings.  In such a description the sequence
consists of coherent sections separated by base pairs, on which
decoherence events take place.
The positions of the decoherence base pairs are chosen at random with
a probability $p$ giving rise to a particular decoherence
configuration. The final results will be averaged over the different
decoherence configurations. The average decoherence length is $a/p$ \cite{ZUW_2009} ,
where $a\approx 3.4 \AA $ is the distance between base pairs. $p$ is
the first of the two parameters characterizing decoherence in the
model. 

The coupling of a base pair to the environment implies that the local
energy levels of the two bases will be broadened. In our model this
broadening will be described by a parameter $\eta$.  As we furthermore
assume that coherence is completely destroyed on these base pairs, it
makes sense to attribute a local energy distribution function to each
decoherence base pair, which in general will differ from the Fermi
distribution. Instead it will be a nonequilibrium distribution that
emerges from the Fermi distributions $f_L(E)$ and $f_R(E)$ in the
left, respectively the right contact via the transmission of charge
carriers through the coherent sections.  We expect $\eta$ to be larger
than the broadening $\sim\Gamma/2$ due to the electrodes and of the
same order of magnitude as the width of the subbands in Fig.~\ref{inf_DOS}
($\approx 0.04$eV). 

The $(n-1)$st and the $n$th decoherence base pair can be regarded as
``contacts'' for the coherent section in between. The electrical
current (and hence the conductance) can then be calculated according
to the Landauer formula from the transmission function $T_{n-1,n}(E)$
of the coherent section and the energy distribution functions
$f_{n-1}(E)$ and $f_n(E)$ at its boundaries,
\begin{equation}
I=2\frac{e}{h}\int dE\ T_{n-1,n}(E)\left(f_{n-1}(E)-f_{n}(E)\right).
\label{Landauer}
\end{equation}

\begin{figure}
\includegraphics[width=8truecm]{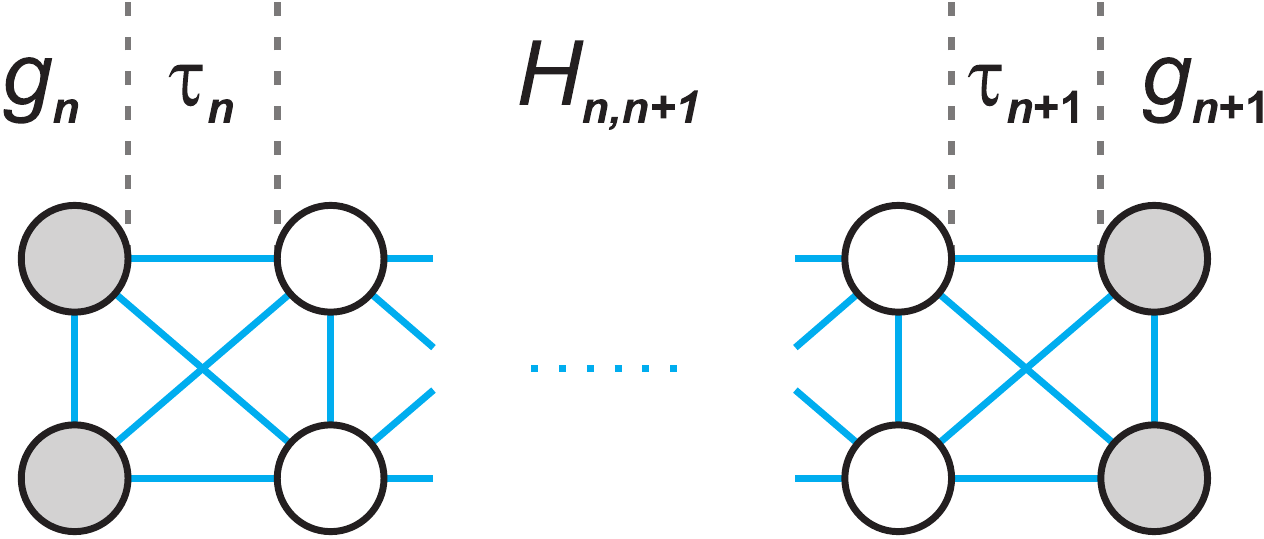}
\caption{The transmission function of a coherent double strand
  DNA molecule segment is calculated from the Hamiltionian given 
  in Section \ref{DNA} and the self-energies describing the 
  connection to the decoherence base pairs at its ends (shaded).
\label{segment}}
\end{figure}

The transmission functions for the coherent sections are calculated
by applying the non-equilibrium Green's function method \cite{Datta_I,*Datta_II} 
(see Fig.~\ref{segment}):
\begin{eqnarray}
\label{transmission}
&&T_{n,n+1}(E)=\cr 4&&\mathrm{Tr}[G_{n,n+1}(E)\Im\Sigma_{n+1}(E)
G_{n,n+1}^\dagger (E)\Im\Sigma_{n}(E)],
\end{eqnarray}
where 
\begin{equation}
  \label{G}
G_{n,n+1}(E)=\left[E-H_{n,n+1}-\Sigma_n(E)-\Sigma_{n+1}(E)\right]^{-1}
\end{equation}
is the Green's function of the segment between decoherence base pairs
$n$ and $n+1$. It is a matrix of dimension $2j_n\times 2j_n$, where $j_n$
denotes the number of base pairs in the segment.  The self-energy is a
product of matrices  
\begin{equation}
\begin{array}{ccccc}
\Sigma_n(E)         &=&   \qquad   \tau_n       &   \qquad  g_n(E)   &\qquad\tau^\dagger_n\\
(2j_n\times 2j_n)  &  &\qquad (2j_n\times  2) &\qquad (2\times 2)&\qquad (2\times 2j_n)
\end{array}
\end{equation}
with the dimensions given in the second
line. $\tau_n$ is the coupling matrix connecting the segment to the 
decoherence base pair $n$ (see Fig.~\ref{segment}). Its matrix
elements can be identified from Fig.~\ref{DNA_ext_model}. 
The Green's function of the decoherence base pair $n$ contains the
corresponding Hamiltonian $H_n$ and the coupling to the environment
described in the simplest case by a constant imaginary self-energy $-\mathrm{i}\eta$
(times the $2\times 2$ unit matrix):
\begin{equation}
g_n(E)=[E-H_{n}+\mathrm{i}\eta]^{-1}.
\end{equation}

As we assume complete loss of phase information at the decoherence
regions, one can write down master equations for the distribution
functions, where the transfer rate 
of charge carriers between neighboring decoherence regions 
is proportional to the transmission function \cite{ZUW_2009}.
In the stationary state one obtains the following system of
coupled linear equations:
\begin{equation}
T_{n-1,n}(E) [f_{n-1}(E)-f_n(E)]  =  T_{n,n+1}(E) [f_n(E)-f_{n+1}(E)].
\label{master_equation}
\end{equation}
According to Eq.~(\ref{Landauer}), equations (\ref{master_equation})
imply current conservation. They can be solved analytically
\cite{ZUW_2009}, using the boundary conditions 
$f_0(E)=f_L(E)$ and $f_{M+1}(E)=f_R(E)$, where $M$ denotes the
number of decoherence base pairs. Inserting the solution into
Eq.~(\ref{Landauer})  (e.g. for $n=M$) gives the current
\begin{equation}
\label{current_solution}
I=2\frac{e}{h}\int dE\,\left(\sum\limits_{n=0}^{M}
\cfrac{1}{T_{n,n+1}}\right) ^{-1} (f_L(E)-f_{R}(E)).
\end{equation}
This leads to the linear conductance
\begin{equation}
\label{linear_conductance}
G=G_0\int dE\,\left(\sum\limits_{n=0}^{M}
\cfrac{1}{T_{n,n+1}}\right) ^{-1} \left(-\left.\frac{\partial f_{\rm
      eq}}{\partial E}\right|_{\mu}\right),
\end{equation}
where $G_0=2\frac{e^2}{h}$ is the conductance quantum (for the two
spin channels), $f_{\rm eq}$ is the Fermi function and $\mu$ is the equilibrium
chemical potential of the contacts. For the results presented in the
following, this expression has been averaged over typically 2000
independent decoherence configurations.

It is worth noting that temperature enters the result via several 
parameters. First, there is the explicite temperature dependence of
the Fermi function. Moreover, the chemical potential $\mu$ and the 
decoherence parameters $p$ and $\eta$ implicitely depend on
temperature.  

Given the microscopic energy values in Tables
~\ref{DNA_on-site_intrabasepair} and \ref{DNA_parameters} and the
temperature, the model has four 
parameters. Two of them, $\mu$ and $\Gamma$, describe the coupling to
the contacts. The other two are the decoherence parameters, $p$ and
$\eta$. As they describe a coupling to the environment, they may
differ for dry and wet DNA, and they may also depend on the type of
base pair.

\section{Results   
\label{exp}}

In this Section we present our results on the linear conductance for the sequences
investigated experimentally in 
\cite{Xu-Zhang-Li-Tao_2004,Hihath-Xu-Zhang-Tao_2005,Hihath-Chen-Zhang-Tao_2007}.
The experimental data consist of conductance values for
5'-(CG)$_{m}$-3' samples \cite{Xu-Zhang-Li-Tao_2004,
Hihath-Chen-Zhang-Tao_2007} with four different lengths (numbers of base
pairs (nBP) are $2m=8,10,12,14$),
and for 5'-CGCG-(A)$_n$(T)$_n$-CGCG-3' samples \cite{Xu-Zhang-Li-Tao_2004,
Hihath-Xu-Zhang-Tao_2005} with three different
lengths (nBPs are $8+2n=8,10,12$), where $n=0$ corresponds to the
shortest 5'-(CG)$_{m=4}$-3' sample. 

We took the effect of finite temperature into account only in the
Fermi functions of the electrodes. Since the experiments were
performed at room temperature, we fixed $k_BT=0.0255$eV. It is
interesting to note, that in \cite{Hihath-Chen-Zhang-Tao_2007}
experimentally unobservable temperature dependence was reported 
in the temperature range from 5 to 65 Celsius.  

It is a hard task to investigate the whole  four dimensional
parameter space of the model presented in Sect.~\ref{decoh}. One needs
to have reasonable guesses for the parameter values $\Gamma,\ \mu,\ p,
$ and $\eta$. It turns out that the chemical potential $\mu$ of the
electrodes is the most crucial parameter. Here we present data for four
values, $\mu = 8.07$eV ($\mu_{\rm el}$ slightly above the upper band),
$\mu = 8.18$eV ($\mu_{\rm el}$ within the upper band), and $\mu =
8.25$eV, respectively $\mu = 8.26$eV (both $\mu_{\rm el}$ slightly
below the upper band of the isolated molecule). The coupling $\Gamma$
to the electrodes is assumed to be of the same order of magnitude as
the weakest hopping parameters in the DNA molecule. It is fixed to
the value $\Gamma=0.003$eV in all figures apart from the last one. As
explained in Sec.\ref{decoh}, we regard values of $\eta$ 
slightly larger than 0.04eV (but still of the same order of magnitude) 
as reasonable. Finally, we guess that at room temperature a hole
cannot travel ballistically further than perhaps two base pairs. Thus
we choose $p=0.5$. These guesses will be confirmed by the results.

\begin{figure}
\includegraphics[width=8.5truecm]{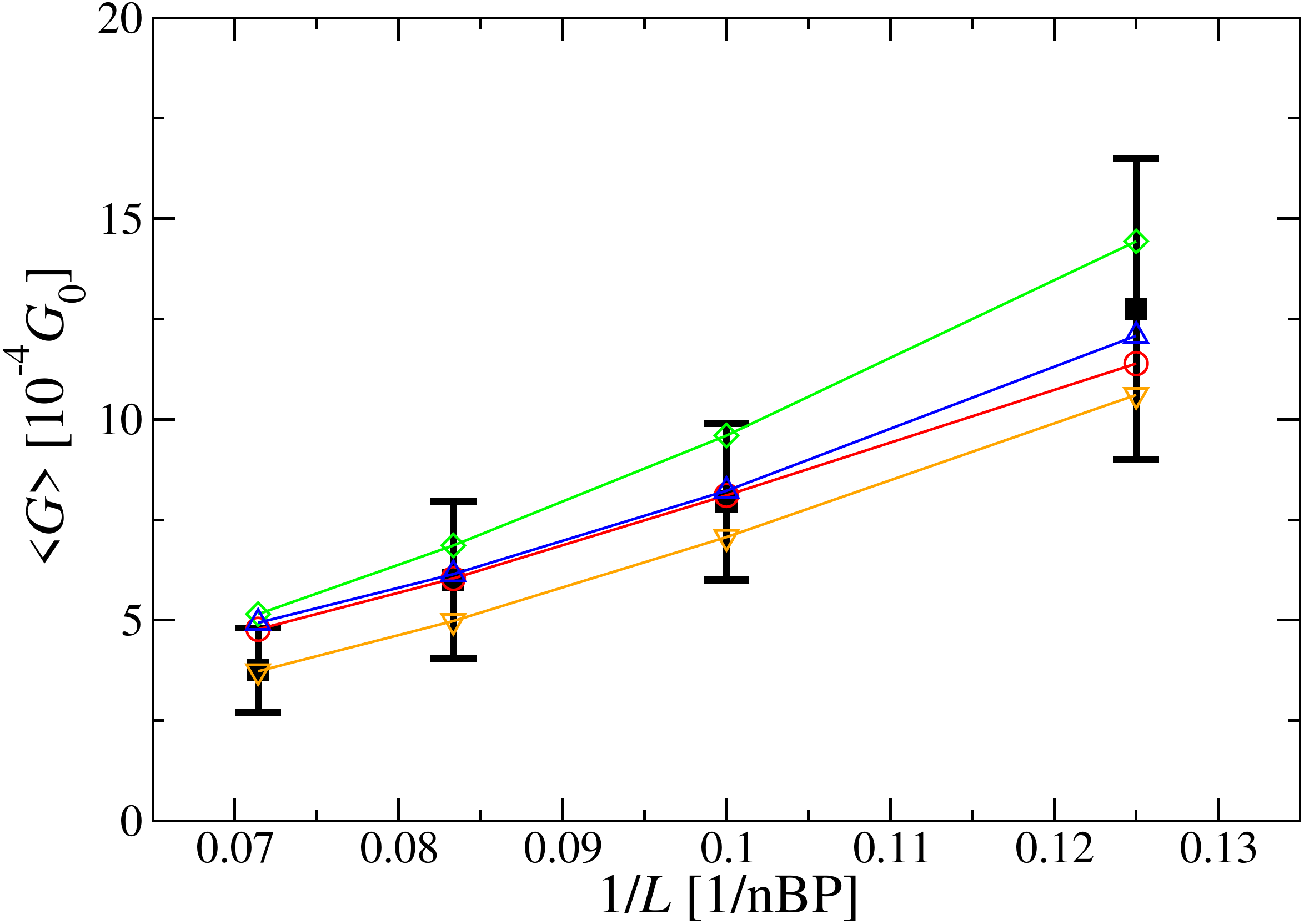}
\caption{Calculated conductances for 
5'-(CG)$_{m}$-3' versus the inverse of the number of base
pairs (nBP) $2m$ for parameter
combinations ($\mu=8.07$eV, $\eta=0.05$eV) (red), ($\mu=8.18$eV,
$\eta=1.0$eV) (blue),
($\mu=8.25$eV, $\eta=0.05$eV) (green),
and ($\mu=8.26$eV, $\eta=0.05$eV) (orange). 
$\Gamma=0.003$eV, $p=0.5$ and $k_B T=0.0255$eV are the
same for all curves. The experimental values \cite{Xu-Zhang-Li-Tao_2004,
Hihath-Chen-Zhang-Tao_2007} are shown by $\blacksquare$ together with
the error bars.
\label{linear_G_vs_inverse_length}}
\end{figure}

Fig.~\ref{linear_G_vs_inverse_length} shows the measured conductances
of four molecules of the type 5'-(CG)$_{m}$-3', plotted versus the
inverse of their length $L = 2m$ (number of base pairs (nBP))
\cite{Xu-Zhang-Li-Tao_2004, Hihath-Chen-Zhang-Tao_2007}. Within the
experimental error bars we obtained four excellent fits by keeping
$p=0.5$ and $\Gamma=0.003$eV fixed and optimizing $\eta$ for each
value of the chemical potential. The statistical errors of the
calculated conductances are not shown in this 
figure for the sake of clarity. They are always less than half of the
experimental errors. In the cases, where the chemical potential lies
outside the band, the value $\eta=0.05$eV led to good fits for all
four data points. However, if the chemical potential lies inside the
band, a good fit requires a value of $\eta=1$eV, which seems
unreasonably large.   

\begin{figure}
\includegraphics[width=8.5truecm]{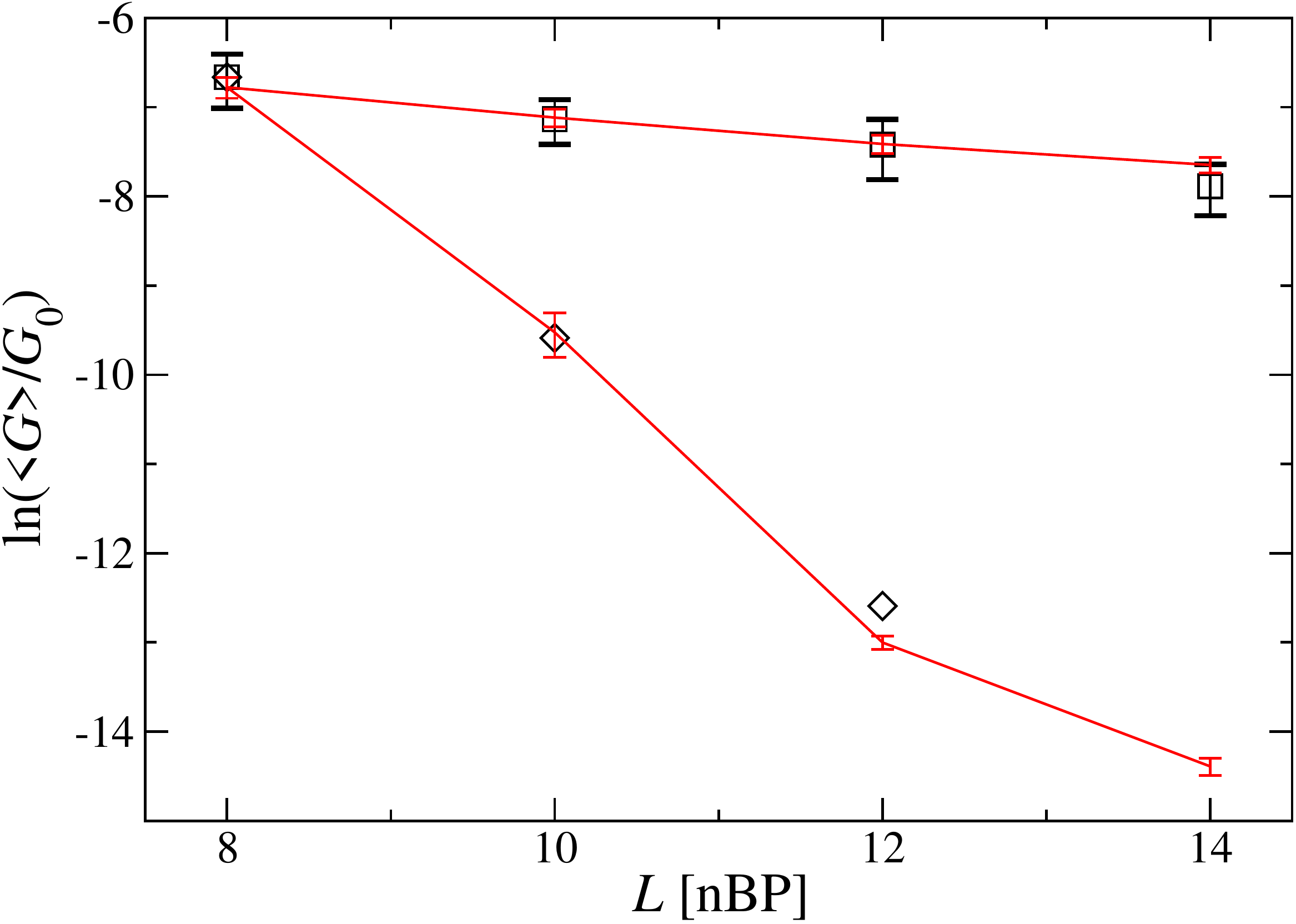}
\caption{Conductance data for 
5'-(CG)$_{m}$-3' ($\Box$) \cite{Xu-Zhang-Li-Tao_2004,
Hihath-Chen-Zhang-Tao_2007}
and  
5'-CGCG-(A)$_n$(T)$_n$-CGCG-3' ($\Diamond$) \cite{Xu-Zhang-Li-Tao_2004,
Hihath-Xu-Zhang-Tao_2005}
plotted semilogarithmically versus the number of base pairs (nBP),
$2m$, respectively $8+2n$. 
For a value of $\mu=8.07$eV, our model fits both data sets, using the
same parameters
$\Gamma=0.003$eV, $p=0.5$, and $\eta=0.05$eV. $k_B T=0.0255$eV. 
\label{semilog_G_vs_length_1}}
\end{figure}

\begin{figure}
\includegraphics[width=8.5truecm]{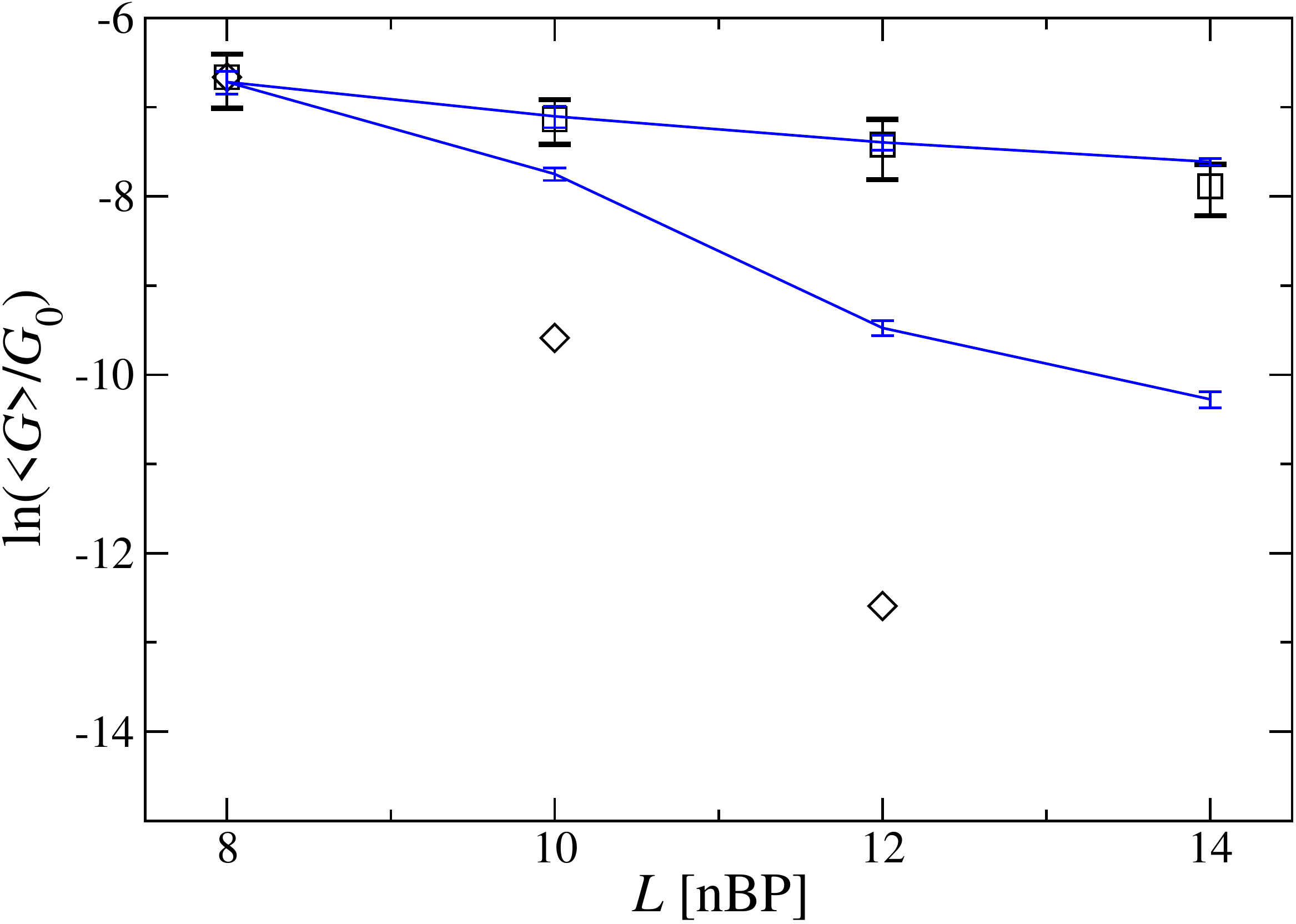}
\caption{Same experimental data as in Fig.~\ref{semilog_G_vs_length_1}. 
For a value of $\mu=8.18$eV, the parameter
  values fitting 5'-(CG)$_{m}$-3', $\Gamma=0.003$eV, $p=0.5$, $\eta=1.0$eV, do
  not fit the data for the chains with AT-inclusions. $k_B T=0.0255$eV.
\label{semilog_G_vs_length_2}}
\end{figure}

\begin{figure}
\includegraphics[width=8.5truecm]{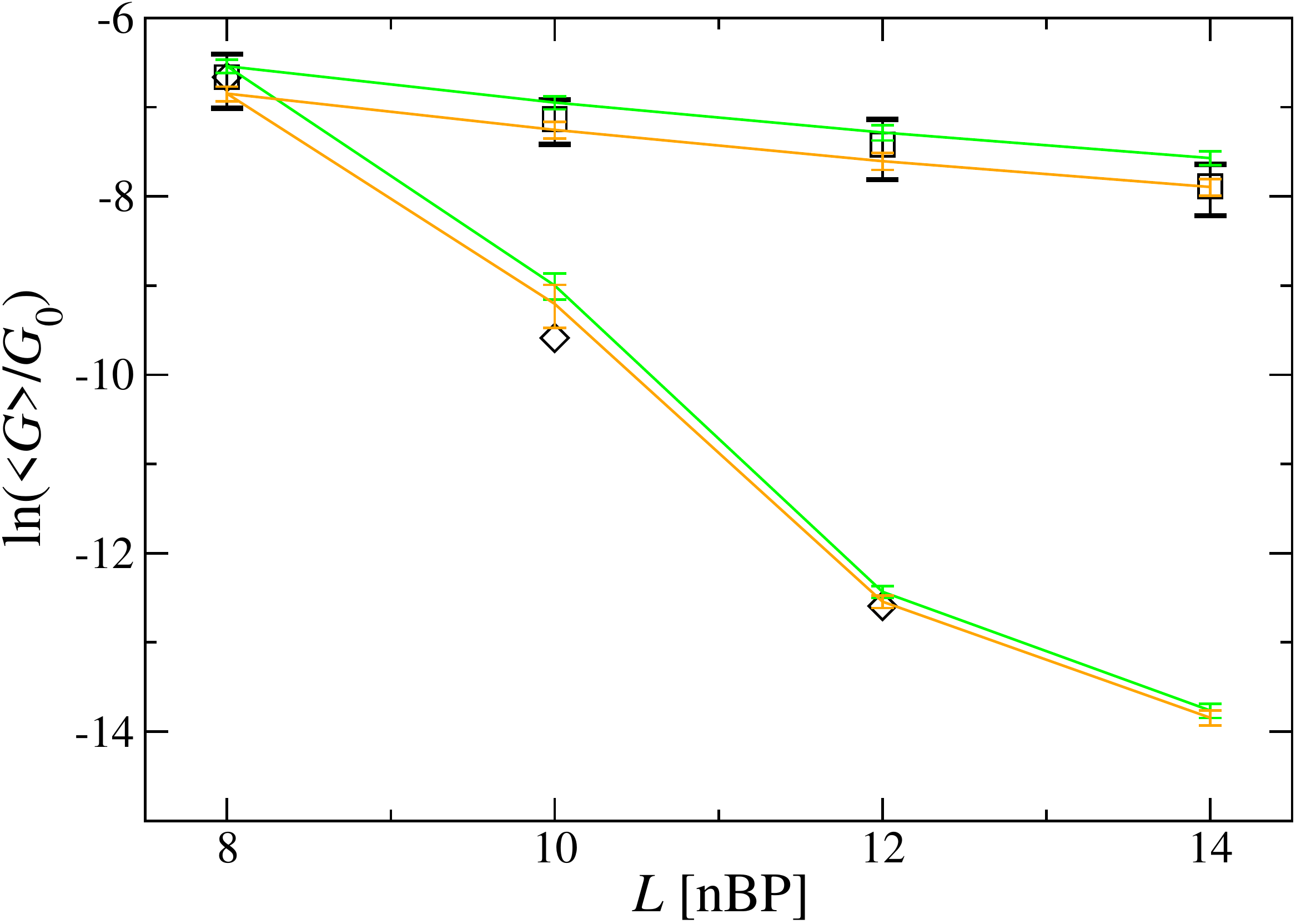}
\caption{Same experimental data as in Fig.~\ref{semilog_G_vs_length_1}. 
For values between
  $\mu=8.25$eV (green) and $\mu=8.26$eV (orange), our model fits both data
  sets using the same parameters $\Gamma=0.003$eV, $p=0.5$, $\eta=0.05$eV. 
  $k_B T=0.0255$eV.
\label{semilog_G_vs_length_3}}
\end{figure}

Plotting the conductance as a function of inverse length suggests
ohmic behavior. This is misleading, though, because a linear
extrapolation would lead to vanishing conductance at a finite length. 
In Figs.~\ref{semilog_G_vs_length_1},~\ref{semilog_G_vs_length_2},
and~\ref{semilog_G_vs_length_3} we plot the same data in a
semilogarithmic way, showing that an exponential length dependence
cannot be ruled out.

These Figures also show two more experimental data points, which
belong to molecules 5'-CGCG-(A)$_n$(T)$_n$-CGCG-3' with $n=1$ and
$n=2$. We calculated the corresponding conductances using the same
parameters as obtained in Fig.~\ref{linear_G_vs_inverse_length} for
the other molecules. Only the on-site energies and hopping parameters
were adapted according to the Tables~\ref{DNA_on-site_intrabasepair}
and \ref{DNA_parameters}. For the cases $\mu=8.07$eV, $\mu=8.25$eV,
respectively $\mu=8.26$eV excellent fits for the two new data points
were obtained (see Figs.~\ref{semilog_G_vs_length_1} and
~\ref{semilog_G_vs_length_3}). However, for $\mu=8.18$eV, the new
data points cannot be fitted with the old parameters. This casts
further doubt on the validity of this parameter set, in addition to
$\eta$ being unreasonably large.

In the following we examine, how sensitively the calculated
  conductance for the longest of the 5'-(CG)$_{m}$-3' samples ($m=7$)
  depends on the parameters $\eta$, $p$, and $\Gamma$.  The purpose is
  to show that the values leading to the fits in
  Fig.~\ref{linear_G_vs_inverse_length} do not offer much freedom of
  choice, when varied individually.  The chemical potential is fixed
  at $\mu=8.07$eV, $\mu=8.18$eV, respectively $\mu=8.25$eV,
  corresponding to the fit curves close to the upper end of the error
  bar for this sample. It should be kept in mind, that a change of
  parameters that moves this conductance closer to the experimental value,
  will in general lead to worse fits for the other three samples in 
  Fig.~\ref{linear_G_vs_inverse_length}.

\begin{figure}
\includegraphics[width=8.5truecm]{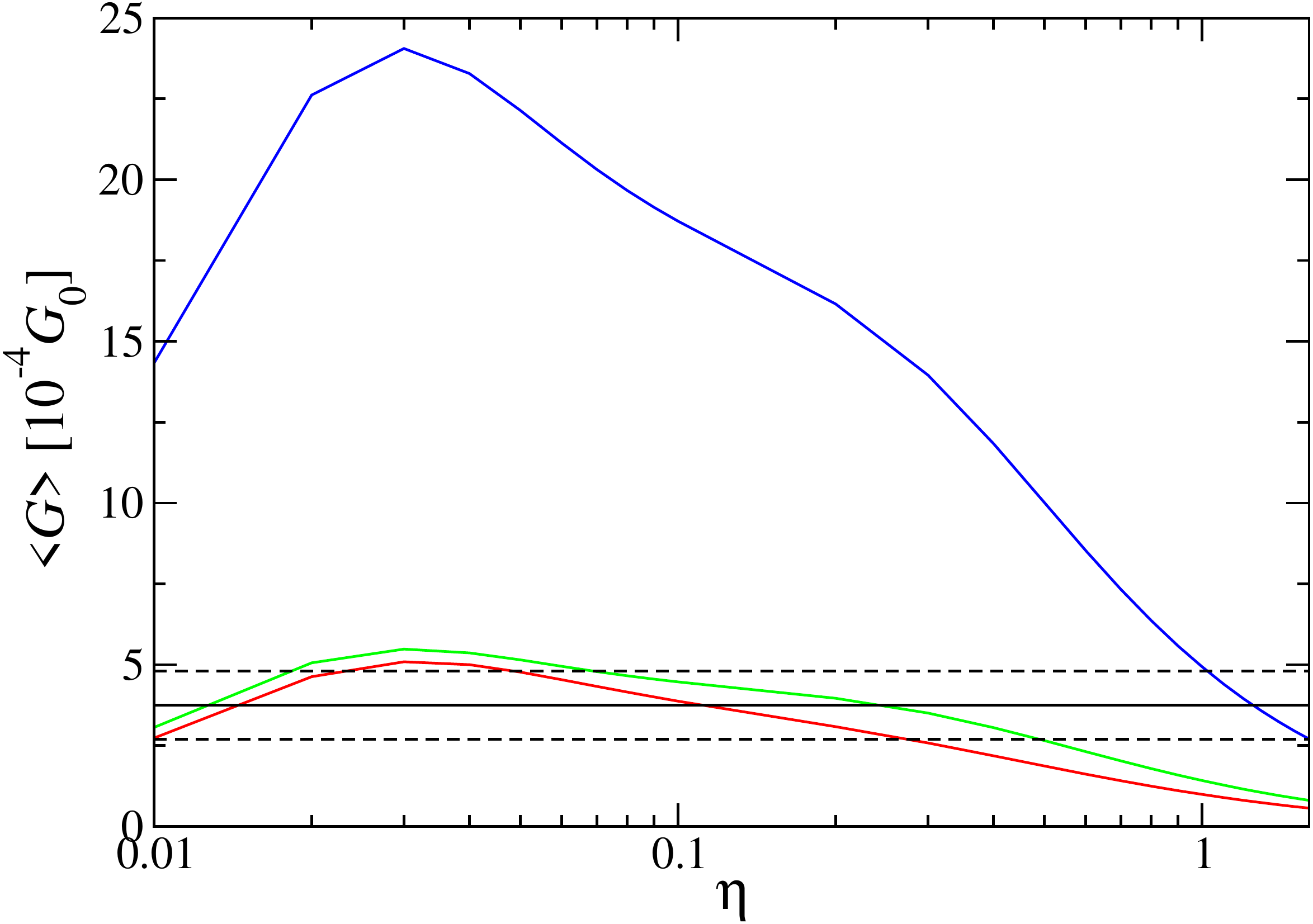}
\caption{The sensitivity of the calculated conductance on $\eta$.
The functions $G(\eta)$ are shown for $\mu=8.07$eV (red), $\mu=8.18$eV
(blue), and $\mu=8.25$eV (green) for
5'-(CG)$_{7}$-3'. $\Gamma=0.003$eV and
  $p=0.5$ are fixed. The experimental value
  is indicated by the horizontal full line, its errorbar by the
  dashed ones. $k_B T=0.0255$eV as
  in the experiment. 
\label{eta_fit_vs_mu}}
\end{figure}

In Fig.~\ref{eta_fit_vs_mu} the calculated conductance is plotted as
a function of $\eta$ while keeping $p=0.5$ and 
$\Gamma=0.003$eV fixed on the values, for which the fits 
in Fig.~\ref{linear_G_vs_inverse_length} were obtained.
For $\mu=8.07$eV (red) and $\mu=8.25$eV (green) the conductance does
not depend sensitively on $\eta$: One would get acceptable fits for
values in a large interval between 0.01eV and 0.5eV. This is true for a
molecule that is 14 base pairs long. The optimization of the
$\eta$-value has to rely on the length dependence of the
conductance. These data are crucial to narrow down the interval for
this model parameter. By contrast, for $\mu=8.18$eV the conductance
depends very sensitively on $\eta$, allowing acceptable fits only in a
narrow interval around 1eV and perhaps around a second value much below
0.01eV. Both values are outside the range, which we regard as plausible.

\begin{figure}
\includegraphics[width=8.5truecm]{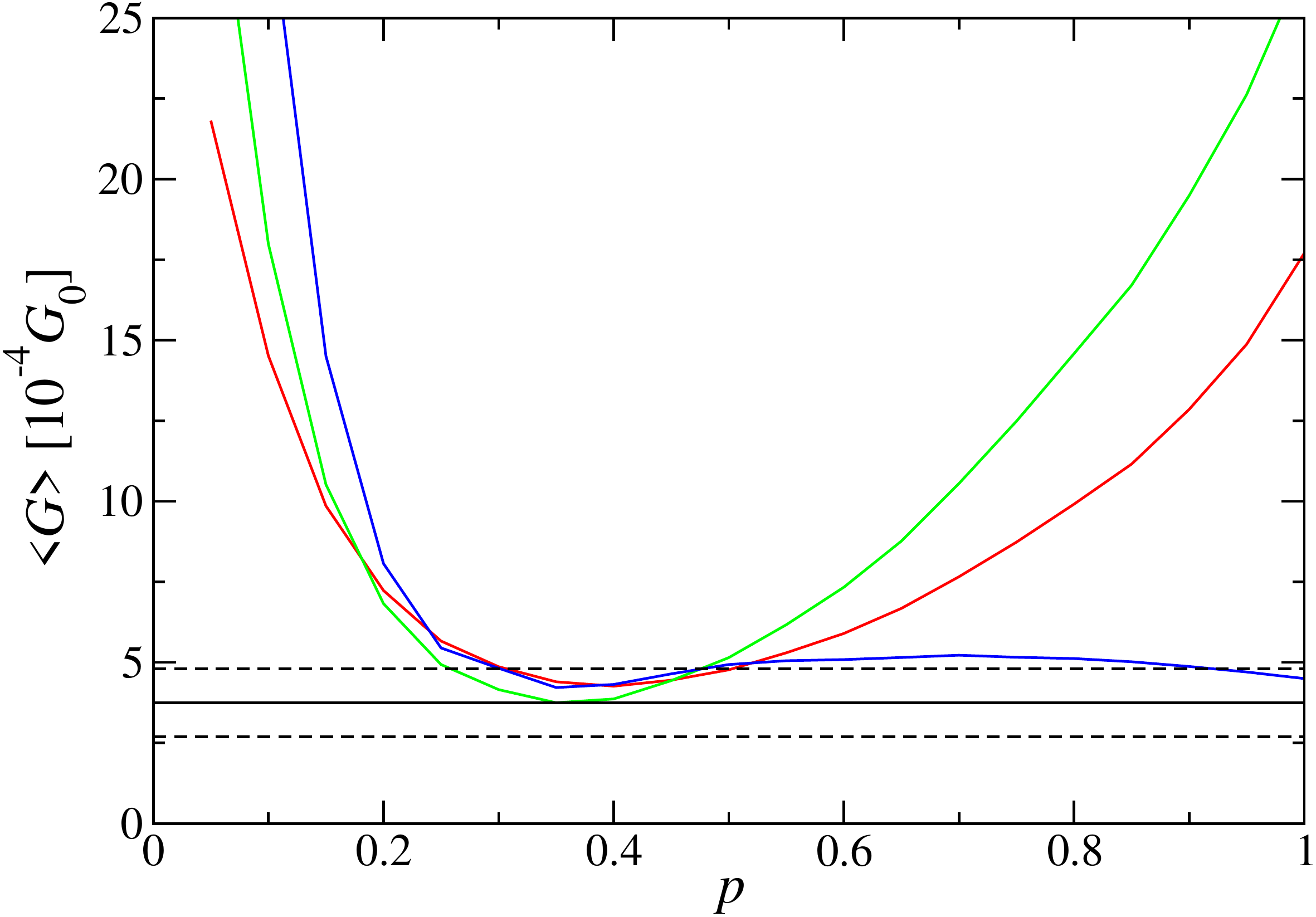}
\caption{The sensitivity of the calculated conductance on $p$. 
The functions $G(p)$ are shown for $\mu=8.07$eV, $\eta=0.05$eV (red), 
  $\mu=8.18$eV, $\eta=1.0$eV (blue), and $\mu=8.25$eV, $\eta=0.05$eV (green) 
  for 5'-(CG)$_{7}$-3'. $\Gamma=0.003$eV. The experimental value
  is indicated by the horizontal full line, its errorbar by the
  dashed ones. $k_B T=0.0255$eV as
  in the experiment. 
\label{p_fit_vs_mu}}
\end{figure}

In Fig.~\ref{p_fit_vs_mu} the calculated conductance is plotted as
a function of $p$ while keeping $\eta=0.05$eV respectively $\eta=1$eV
and $\Gamma=0.003$eV fixed on the values giving the fits 
in Fig.~\ref{linear_G_vs_inverse_length}. For the chemical potentials
outside the band ($\mu=8.07$eV (red) and $\mu=8.25$eV (green)) the
sensitivity is high: Only for $p$-values in the interval between 0.3
and 0.5 one gets acceptable fits. However, for the chemical potential
within the band ($\mu=8.18$eV) the sensitivity is low: Considering
only the sample with 14 base pairs, any $p$-value between 0.3 and 1
would be allowed.

\begin{figure}
\includegraphics[width=8.5truecm]{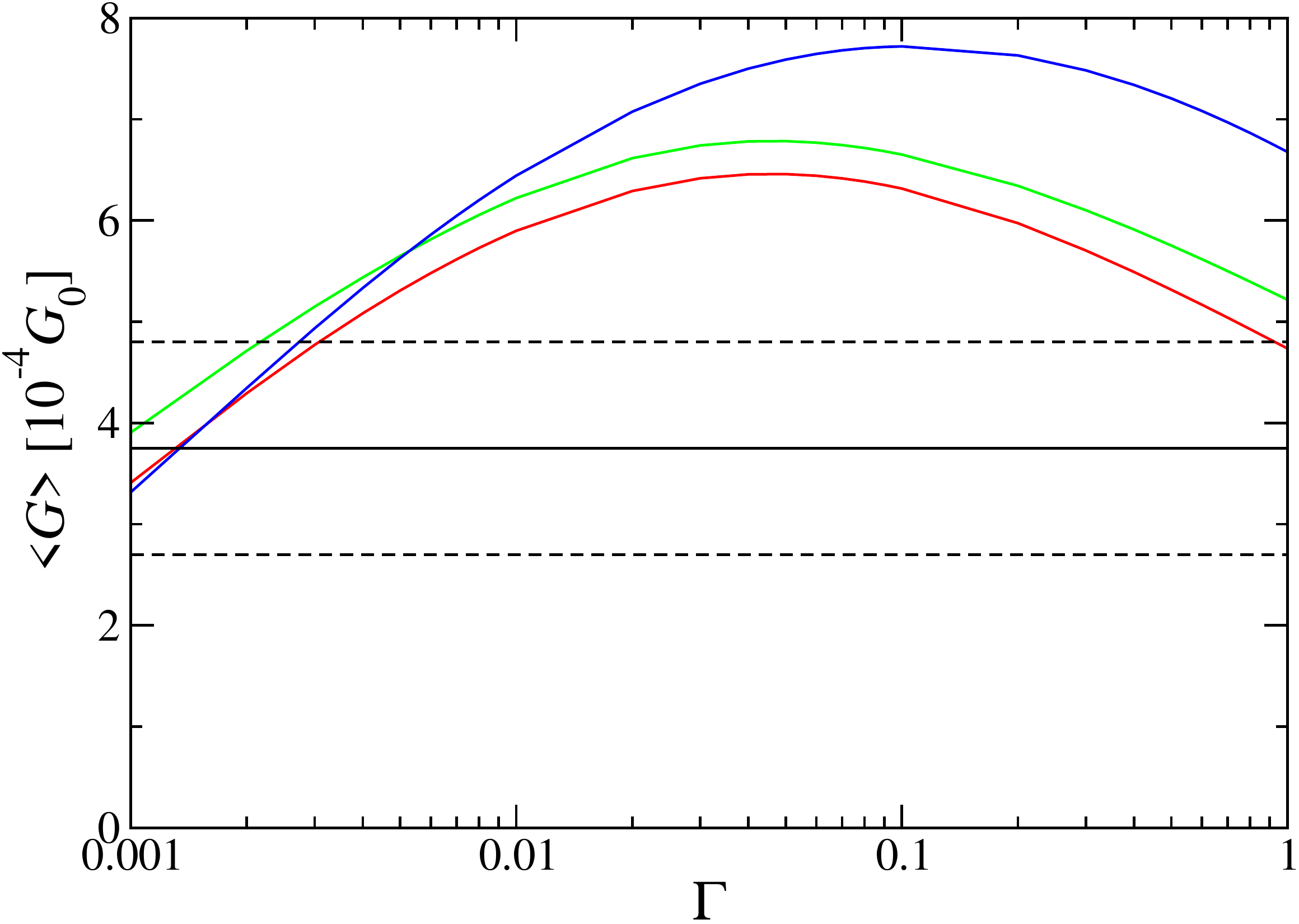}
\caption{The sensitivity of the calculated conductance on $\Gamma$.
 The functions $G(\Gamma)$ are shown for $\mu=8.07$eV, $\eta=0.05$eV (red), 
  $\mu=8.18$eV, $\eta=1.0$eV (blue), and $\mu=8.25$eV, $\eta=0.05$eV (green) 
  for 5'-(CG)$_{7}$-3'. $p=0.5$. The experimental value
  is indicated by the horizontal full line, its errorbar by the
  dashed ones. $k_B T=0.0255$eV as in the experiment. 
\label{Gamma_fit_vs_mu}}
\end{figure}

In Fig.~\ref{Gamma_fit_vs_mu} the calculated conductance is plotted as
a function of  $\Gamma$ while keeping $p=0.5$ and $\eta$ fixed on the values
giving the fits in Fig.~\ref{linear_G_vs_inverse_length}. For all
three chemical potentials the conductance depends sensitively on
$\Gamma$. Acceptable fits are obtained between values somewhat below
0.001eV and 0.003eV, justifying our weak coupling assumption. There
may be a second fit interval with very large values above 1eV,
which we discard as unplausible, because they would imply a
coupling of the molecule to the electrodes that is better than any
coupling between the base pairs within the molecule.

\section{Conclusions and Outlook\label{concl}}

We have proposed a simple model in order to calculate the conductance
of DNA-double strand molecules for any sequence of base pairs. It is a
tight binding model with on-site energies and charge transfer
integrals taken from the DFT calculations in \cite{Senthil_2005}. 

The on-site
energies correspond to the energy difference between the vacuum level
and the highest occupied molecular orbital (HOMO) of the nucleobase.
In the 5'-(CG)$_{\infty}$-3' double strand they give rise to a lower band
about 9.72eV, and an upper band about 8.18eV below the vacuum level.
Both of them are filled, if the molecule is neutral. The energy to
create an electron-hole pair in such a molecule is of the order of 
$\min(\text{LUMO}_\text{C},\text{LUMO}_\text{G}) + 8.18\text{eV} \approx 3-4$eV
\cite{*[{For a summary, see }] Endres-Cox-Singh_2004}, 
where $\text{LUMO}_\text{C/G}$ denotes the energy of the
lowest unoccupied molecular orbital of C respectively G. Actually the
difference between the band centers overestimates the band gap, but as
the band width is of the order of 0.05eV, the conclusion is still
correct that thermal activation of charge carriers at room temperature
can be ruled out.

The situation changes, if the DNA-molecule is brought into contact
with gold electrodes. Let us first discuss the case, where the 
chemical potential of holes in the electrode is $\mu=8.07$eV (in
other words, electronic states are filled up to the energy $-8.07$eV). At room
temperature there will be thermally activated holes in the electrode 
down to the HOMO$_\text{G}$ level. They can be injected into the
molecule and carry a current through the molecule. In this sense the
electrode has a similar effect as acceptors in a p-doped semiconductor.
This works only, if the molecule is short enough that no localized
space charges form at the contacts. In summary, for $\mu=8.07$eV the
molecule is a hole conductor, which remains neutral when contacted.

If the chemical potential of holes in the electrode is $\mu = 8.26$eV, we obtained fits of the experimental data of 
similar quality. In this case, however, the Fermi energy of the gold
electrodes lies below the uppermost filled band of the DNA molecule.
Again we assume that the molecule is short enough that space charges
extend all the way from one side to the other. Then the upper band
will be depleted, the molecule will carry a positive charge of $2N$
elementary charges, where $N$ is the number of base pairs. In the
electrodes there will be thermally activated 
electrons reaching up to the now empty HOMO$_\text{G}$ level. In this
case the electrodes act like donors in an n-doped semiconductor.
In summary, for $\mu=8.26$eV the molecule would become charged when
contacted, and it would be an electron conductor.

The third parameter set, for which we could find an excellent fit of
the conductance of 5'-(CG)$_{n}$-3' double strands, had $\mu=8.18$eV,
which lies within the upper occupied band of the DNA molecule. This
will lead to a depletion of the band, which is only partial, so that no thermal
activation of charge carriers is needed. In this sense the DNA
molecule behaves like a metal, provided it is shorter than the
screening length. 
However, the fit requires a decoherence induced
level broadening $\eta=1$eV, which is unphysically large. Moreover,
the same parameter set is incompatible with the conductance data for
the mixed 5'-CGCG-(A)$_n$-(T)$_n$-CGCG-3' double strands. Therefore,
this scenario can be ruled out. 

Most of the experimental and the theoretical
examinations refer to hole conduction. Therefore we can 
conclude that the first parameter set ($\mu=8.07$eV, $\eta=0.05$eV,
$\Gamma = 0.003$eV, and $p = 0.5$) is the most plausible one to describe
the DNA molecules considered.

Summarizing, we extended our phenomenological model
\cite{ZUW_2009,MatiasPhD} for describing the
effect of decoherence in double stranded DNA molecules. We could fit
all of the experimental data of  
\cite{Xu-Zhang-Li-Tao_2004,Hihath-Xu-Zhang-Tao_2005,Hihath-Chen-Zhang-Tao_2007}
without changing the microscopic energy values and by using the same
parameter set of the model.
For other experimentally investigated sequences
\cite{Mahapatro-Jeong-Lee-Janes_2007,Dulic-Tuukkanen-Chung-Isambert-Lavie-Filoramo_2009}
work is in progress. 

\begin{acknowledgments}
We would like to thank NJ Tao for helpful information on the details
of their experiments. This work was supported by DFG-grant GRK 1240
``Nanotronics''. O.U. acknowledges the support of the Alexander von
Humboldt Foundation and the J\'anos Bolyai Research Scholarship of the 
Hungarian Academy of Sciences.
\end{acknowledgments}

\bibliography{DNA}

\end{document}